\documentclass[12pt,a4paper]{article}
\usepackage{xcolor}
\usepackage{jheppub}
\usepackage{epstopdf}
\usepackage{graphicx}
\usepackage{epsfig}
\usepackage{dcolumn}  
\usepackage{bm}    
\usepackage{amssymb} 
\usepackage{amsmath,bm}
\usepackage{amsfonts}    
\usepackage{slashed}  
\usepackage{youngtab}
\usepackage[mathscr]{euscript}
\usepackage{epsfig}
\usepackage{verbatim}
\newdimen\tableauside\tableauside=1.0ex
\newdimen\tableaurule\tableaurule=0.4pt
\newdimen\tableaustep
\def\phantomhrule#1{\hbox{\vbox to0pt{\hrule height\tableaurule width#1\vss}}}
\def\phantomvrule#1{\vbox{\hbox to0pt{\vrule width\tableaurule height#1\hss}}}
\def\sqr{\vbox{%
		\phantomhrule\tableaustep
		\hbox{\phantomvrule\tableaustep\kern\tableaustep\phantomvrule\tableaustep}%
		\hbox{\vbox{\phantomhrule\tableauside}\kern-\tableaurule}}}
\def\squares#1{\hbox{\count0=#1\noindent\loop\sqr
		\advance\count0 by-1 \ifnum\count0>0\repeat}}
\def\tableau#1{\vcenter{\offinterlineskip
		\tableaustep=\tableauside\advance\tableaustep by-\tableaurule
		\kern\normallineskip\hbox
		{\kern\normallineskip\vbox
			{\gettableau#1 0 }%
			\kern\normallineskip\kern\tableaurule}%
		\kern\normallineskip\kern\tableaurule}}
\def\gettableau#1 {\ifnum#1=0\let\next=\null\else
	\squares{#1}\let\next=\gettableau\fi\next}

\allowdisplaybreaks[1]

\newcommand{\be}{ \begin{equation}}
\newcommand{\ee}{\end{equation}}
\newcommand{\bea}[1]{\begin{eqnarray}\label{#1} }
\newcommand{\eea}{\end{eqnarray}}

\def\ZZZ{{\hskip-3pt\hbox{ Z\kern-1.6mm Z}}}
\def\zzz{{\hskip-3pt\hbox{ z\kern-1mm z}}}

\newcommand{\cJ}{{\cal J}}
\newcommand{\cK}{{\cal K}}

\def\one{{\hbox{ 1\kern-.8mm l}}}
\def\zero{{\hbox{ 0\kern-1.5mm 0}}}

\title{Tensionless String Spectra on ${\rm AdS}_3$ }

\author{Matthias R.\ Gaberdiel$^a$ and Rajesh Gopakumar$^b$} 
\affiliation{$^a$ Institut f\"ur Theoretische Physik, ETH Zurich, \\
\hspace*{0.3cm}CH-8093 Z\"urich, Switzerland}
\affiliation{$^b$ International Centre for Theoretical Sciences-TIFR, \\
\hspace*{0.3cm}Survey No. 151, Shivakote, Hesaraghatta Hobli, \\
\hspace*{0.3cm}Bengaluru North, India 560 089}

\emailAdd{\!gaberdiel@itp.phys.ethz.ch,~\!rajesh.gopakumar@icts.res.in}

\abstract{The spectrum of superstrings on ${\rm AdS}_3 \times {\rm S}^3 \times \mathbb{M}_4$ with pure NS-NS flux is analysed for the background where the radius of the AdS space takes the minimal value $(k=1)$. Both for $\mathbb{M}_4={\rm S}^3 \times {\rm S}^1$ and $\mathbb{M}_4 = \mathbb{T}^4$ we show that there is a special set of physical states, coming from the bottom of the spectrally flowed continuous representations, which agree in precise detail with the single particle spectrum of a free symmetric product orbifold. For the case of ${\rm AdS}_3 \times {\rm S}^3 \times \mathbb{T}^4$ this relies on making sense of the world-sheet theory at $k=1$, for which we make a concrete proposal. We also comment on the implications of this striking result.
}

\begin{document}

\maketitle

%make math in all titles bold
\makeatletter
\g@addto@macro\bfseries{\boldmath}
\makeatother
%end code
\section{Introduction}

String theory on ${\rm AdS}_3$ backgrounds has a number of special characteristics which makes this a good laboratory to gain a better understanding of the gauge-string correspondence and, more generally, of properties of string theory on nontrivial backgrounds. In particular, the dual CFTs are $2$-dimensional, and hence strongly constrained by symmetry considerations. Using these tools one can, for instance, show that 
at special points in the moduli space, there is an enhanced global higher spin symmetry \cite{Gaberdiel:2014cha, Gaberdiel:2015mra, Gaberdiel:2015wpo}  (much bigger than the more familiar Vasiliev higher spin symmetry \cite{Vasiliev:2003ev} also seen in higher AdS spacetimes) which has been dubbed the Higher Spin Square (HSS). 
Originally, the HSS was inferred from the single particle spectrum of the dual symmetric product orbifold CFT \cite{Gaberdiel:2015mra, Gaberdiel:2015wpo}.
Among other things, this enabled one to make precise the sense in which the symmetric product orbifold describes a tensionless point in the moduli space of string theory on AdS$_3$ \cite{Gaberdiel:2014cha}. 

Another special feature of string theory on AdS$_3$ is the presence of a complete world-sheet description when the three form flux is pure NS-NS \cite{Maldacena:2000hw, Maldacena:2000kv, Maldacena:2001km}. 
Recently, it was observed, both in the bosonic string theory \cite{Gaberdiel:2017oqg} as well as in the superstring theory \cite{Ferreira:2017pgt} with NS-NS flux, that there is a special minimal radius where the spectrum again exhibits a tensionless behaviour in that there is a tower of massless higher spin states. These massless states arise from the ($w=1$) spectrally flowed continuous representation of the ${\rm sl}(2, \mathbb{R})_k$ WZW model which describes the AdS$_3$ background \cite{Gaberdiel:2017oqg, Ferreira:2017pgt}. The minimal radius is given by $k=3$ for the bosonic case, which translates into $k=1$ for the superstring case. At these values the massless states sit at the bottom of a continuum of states which describe the excitations of a single long string in AdS$_3$. (Note that for $k=1$ the discrete spectrum (coming from the unflowed sector) disappears and merges with the continuum, i.e., the lowest excitations come from the sector with $w=1$.) This continuum is a feature of the pure NS-NS background and is expected to be lifted if one turns on an infinitesimal RR three form flux. 

In this paper, we analyse the full spectrum,  not just the massless states, of this superstring theory at the minimal radius $(k=1)$. We concentrate mainly on the special subsector consisting of the lowest states in the spectrally flowed continuous representations (with $w\geq 1$). 
Rather remarkably, we are able to show that this set of states matches, on the nose, with that of the single-particle spectrum of a symmetric product orbifold theory, where the spectral flow parameter $w \in \mathbb{N}$ can be identified with the length of the twisted cycle in the symmetric product orbifold theory. 

We consider both ${\rm AdS}_3\times {\rm S}^3 \times {\rm S}^3 \times {\rm S}^1$ as well as ${\rm AdS}_3 \times {\rm S}^3 \times \mathbb{T}^4$. It turns out that in the former case the relevant symmetric orbifold is that of two free bosons and eight free fermions, i.e., the symmetric orbifold of the so-called $({\cal S}_0)^2$ theory. This is closely related to the symmetric orbifold of ${\cal S}_0$ that was proposed as the CFT dual of this background in \cite{Eberhardt:2017pty}. The analysis for the case of $\mathbb{T}^4$ is somewhat subtle since the world-sheet theory contains a non-unitary $\mathfrak{su}(2)$ algebra at level $\kappa=-1$. We make a proposal for how to make sense of this theory by describing this $\mathfrak{su}(2)_{-1}$ factor in terms of four symplectic bosons \cite{Goddard:1987td}. With this prescription we then show that the resulting symmetric orbifold is that of four free bosons and four free fermions, i.e., that of ${\mathbb T}^4$. (Since we have not kept track of the momentum states, we cannot distinguish between ${\mathbb R}^4$ or ${\mathbb T}^4$.)

The pure NS-NS background describes a different point in moduli space from the dual symmetric orbifold point, and the detailed agreement of the spectra is therefore quite striking. It indicates, in our opinion, a certain universality about the tensionless limit of AdS$_3$ backgrounds. While both theories seem to be `tensionless', they exhibit not only exactly the same enhanced symmetry (the Higher Spin Square), but also at least some part of the same spectrum.\footnote{The background with NS-NS flux has, however, additional states: in particular we have not just the ground states of the continuum, but also the full continuum. In addition, there are isolated states from the discrete representations, see Section~\ref{sec:2.2}.} This seems to suggest that the presence of the HSS is quite constraining for the full spectrum (which is organised in terms of representations of this symmetry) and not just the massless sector. 

\medskip

The paper is organised as follows. In Section~\ref{sec:worldsheet}, after reviewing some basic facts about the representations which enter in the world-sheet description, we focus on the case of $k=1$, and specifically, on the states that arise from the spectrally flowed continuous representations. We enumerate all physical states that arise in this sector for the case of ${\rm AdS}_3\times {\rm S}^3 \times {\rm S}^3 \times {\rm S}^1$, and determine the corresponding generating function which we then bring into a simple and suggestive form. 
In Section~\ref{sec:symorb} we match this spectrum for the ${\rm AdS}_3\times {\rm S}^3 \times {\rm S}^3 \times {\rm S}^1$ case with the symmetric product orbifold of the $({\cal S}_0)^2$ theory. Section~\ref{sec:symorbT4} discusses the case of the ${\rm AdS}_3 \times {\rm S}^3 \times \mathbb{T}^4$ background. Here we propose a construction of the $k=1$ world-sheet theory, using the free field realisation of $\mathfrak{su}(2)_{-1}$ in terms of four symplectic bosons \cite{Goddard:1987td}, and show that it leads to a sensible interpretation. We find a similar matching of the world-sheet spectrum with that of the spacetime symmetric product orbifold of a $\mathbb{T}^4$ theory. We conclude in Section~\ref{sec:concl} with the discussion of open questions and directions for future research. There is one Appendix in which we have reviewed some aspects of the symmetric orbifold construction.

\section{The World-sheet Spectrum at $k=1$}\label{sec:worldsheet}

Let us begin by reviewing the structure of the world-sheet theory at $k=1$. For the case where ${\rm AdS}_3\times {\rm S}^3$ has pure NS-NS background, the theory is described by a WZW model based on $\mathfrak{sl}(2,\mathbb{R})\oplus \mathfrak{su}(2)$. The bosonic version of this theory was discussed in some detail in the seminal papers \cite{Maldacena:2000hw,Maldacena:2000kv,Maldacena:2001km}; in what follows we will use the supersymmetric version, using the conventions of \cite{Ferreira:2017pgt}, see also \cite{Giveon:1998ns,Israel:2003ry,Raju:2007uj}.

We shall denote the generators of $\mathfrak{sl}(2,\mathbb{R})$ at level $k$ by $J^a_n$ and the associated fermions by $\psi^a_r$, where $a\in\{\pm,3\}$. The fermions transform in the adjoint representation with respect to the generators $J^a_n$, but we can decouple the fermions by introducing the (decoupled) generators $\cJ^a_n$, which then define the affine algebra $\mathfrak{sl}(2,\mathbb{R})$ at level $\kappa=k+2$. The degrees of freedom associated to the ${\rm AdS}_3$ factor contribute therefore
\be
c ( \mathfrak{sl}(2,\mathbb{R}) ) = 3 \Bigl( \frac{k+2}{k} + \frac{1}{2} \Bigr)  
\ee
to the central charge. Similarly, the generators of $\mathfrak{su}(2)$ at level $k'$ are denoted by $K^a_n$ and the associated fermions by $\chi^a_r$, where again $a\in\{\pm,3\}$. The decoupled generators $\cK^a_n$ then define an $\mathfrak{su}(2)$ algebra at level $\kappa'=k'-2$, and the central charge that is associated to the ${\rm S}^3$ factor contributes
\be
c ( \mathfrak{su}(2) ) = 3 \Bigl( \frac{k'-2}{k'} + \frac{1}{2} \Bigr) \  . 
\ee
We shall mainly be interested in two backgrounds. For ${\rm AdS}_3 \times {\rm S}^3 \times \mathbb{T}^4$ the condition that the total central charge equals $c=15$, requires that 
\be
c ( \mathfrak{sl}(2,\mathbb{R}) )  +  c ( \mathfrak{su}(2) ) = 9  \qquad \Rightarrow \qquad k' = k \ . 
\ee
For $k=1$, this leads to $\kappa'=-1$, and hence the world-sheet theory is somewhat singular. In particular, the central charge coming from $\mathfrak{su}(2)$ factor is then negative 
\be\label{csu2}
\left. c ( \mathfrak{su}(2) )  \right|_{k'=1} = - 3  + \frac{3}{2} \ ,
\ee
where $-3$ is the contribution of the decoupled bosonic algebra, while $\frac{3}{2}$ accounts for the three free fermions. We will come back to the question of how to make sense of this world-sheet theory in Section~\ref{sec:symorbT4}.
\smallskip

The other background which we shall discuss, ${\rm AdS}_3 \times {\rm S}^3 \times {\rm S}^3 \times {\rm S}^1$, is actually somewhat simpler. In this case, criticality leads to the condition \cite{Elitzur:1998mm} that 
\be
\frac{1}{k} = \frac{1}{k_+} + \frac{1}{k_-} \ , 
\ee
where $k_\pm$ are the levels of the (supersymmetric) $\mathfrak{su}(2)$ models. In particular, $k=1$ arises then for $k_\pm=2$, for which the decoupled bosonic $\mathfrak{su}(2)$ algebras appear at $\kappa_{\pm}=0$. This simply means that there are no bosonic degrees of freedom associated to the two ${\rm S}^3$'s, and that the only bosonic degrees of freedom that survive are the $3$ excitations from $\mathfrak{sl}(2)$, together with the excitations from the circle theory. On the other hand, there is no reduction in the number of fermionic degrees of freedom. Thus before imposing the physical state condition we have $4$ bosonic and $10$ fermionic world-sheet excitations, which get reduced to $2$ bosonic and $8$ fermionic degrees of freedom after imposing the super Virasoro conditions. 
In the following we shall first treat the case of ${\rm AdS}_3\times {\rm S}^3 \times {\rm S}^3 \times {\rm S}^1$; we will come back to 
${\rm AdS}_3\times {\rm S}^3 \times \mathbb{T}^4$ in Section~\ref{sec:symorbT4}.

\subsection{Discrete and Continuous Representations}

For the case of ${\rm AdS}_3\times {\rm S}^3 \times {\rm S}^3 \times {\rm S}^1$ there are no bosonic excitations from the $\mathfrak{su}(2)$ factors, and we can concentrate on the $\mathfrak{sl}(2,\mathbb{R})$ algebra. As explained in \cite{Maldacena:2000hw}, the relevant representations of  $\mathfrak{sl}(2,\mathbb{R})$ that contribute to the world-sheet spectrum are the discrete and continuous representations, as well as their spectrally flowed images. The discrete representations are labelled by a real number $j$, and have Casimir $C = - j (j-1)$. The no-ghost theorem implies that $j$ has to satisfy the Maldacena-Ooguri (MO) bound \cite{Maldacena:2000hw} which, for $k=1$, takes the form 
\be\label{MObound}
\frac{1}{2} < j  < \frac{k+1}{2} = 1 \ . 
\ee
On the other hand, for the continuous representations $j=\frac{1}{2} + is$, and the Casimir takes the value $C = \frac{1}{4} + s^2$. 

Let us first analyse the standard representations without any spectral flow. For the discrete representations, the mass-shell condition at $k=1$ takes the form 
\be\label{massshell1}
- j (j-1) + N + h_{\rm rest} = \frac{1}{2} \ , 
\ee
where $h_{\rm rest}$ is the conformal dimension coming from the circle theory, and $N$ is the excitation number of all excitations, i.e., of the $4$ bosonic and $10$ fermionic excitations before imposing the super Virasoro constraints. (We are working here in the NS sector; a similar analysis can also be done in the R sector.)

The GSO projection requires that $N$ is half-integer, and hence in particular $N\geq \frac{1}{2}$. It then follows that there can only be physical states if $-j(j-1)\leq 0$. However, the MO bound (\ref{MObound}) implies $-j(j-1)>0$, and thus there are no physical states from the unflowed discrete representations at $k=1$.

For the continuous representations, the analogue of (\ref{massshell1}) is
\be
\frac{1}{4} + s^2 + N + h_{\rm rest} = \frac{1}{2} \ , 
\ee
and the same argument applies. We therefore conclude that no physical states arise from the unflowed sector for $k=1$.

\subsection{Spectrally Flowed Sectors}\label{sec:2.2}

For the flowed discrete representations the mass-shell condition becomes (see eq.~(5.5) of \cite{Ferreira:2017pgt})
\be
- j(j-1) - w \bigl(m + \tfrac{w}{4}\bigr) + N + h_{\rm rest} = \frac{1}{2} \ , 
\ee
where $m$ is the $J^3_0$ eigenvalue and $N$ the excitation number of the state before spectral flow, and the spectral flow parameter takes the values $w\in \mathbb{N}$. (Note that the spectral flow acts on the full supersymmetric algebra, not just the decoupled bosonic algebra.) The GSO projection depends on the cardinality of $w$ as 
\be\label{GSOw}
N + \frac{w+1}{2}  \in \mathbb{N} \ , 
\ee
see the discussion in eq.~(5.7) of \cite{Ferreira:2017pgt}.

Given that there are no physical states from the unflowed discrete representations, one may wonder whether there are any physical states from the flowed discrete representations. It is not difficult to confirm that generically there are: for example, a solution is given by 
\be
w =1 \ , \qquad N=1 \ , \qquad m = j = \sqrt{\frac{1}{4} +h_{\rm rest}} \ ,
\ee
where $h_{\rm rest}$ is taken to satisfy $0<h_{\rm rest}<\frac{3}{4}$ (so that $j$ satisfies the MO bound eq.~(\ref{MObound})). An example with $h_{\rm rest}=0$ is 
\be
w=4 \ , \qquad N=\frac{13}{2} \ , \qquad m = j = \frac{\sqrt{17}-3}{2} \cong 0.56 \ . 
\ee
While there are such states, they will not be the focus of attention in this paper. Instead we shall concentrate on the states that arise from the spectrally flowed continuous representations. 

\subsubsection{Spectrally Flowed Continuous Representations}

For the spectrally flowed continuous representations the mass-shell condition (in the NS sector) takes the form
\be\label{massshellflow}
\frac{1}{4} + s^2 - w  \bigl(m + \tfrac{w}{4}\bigr) + N + h_{\rm rest} = \frac{1}{2} \ .
\ee
There is a similar relation for the right-movers, but with the same value for $w\in\mathbb{N}$. Here $N$ (and similarly $\bar{N}$ for the right-movers) is, as before, the excitation number, and $m$ the $J^3_0$ eigenvalue before spectral flow. The $J^3_0$ eigenvalue, after spectral flow, is then\footnote{We use the symbol $h$ here since this is also the spacetime conformal dimension in the dual CFT. Hopefully this will not be confused with world-sheet conformal dimensions such as $h_{\rm rest}$.}
\be\label{3.2}
h = m + \frac{k}{2} w   =  m + \frac{w}{2} \ . 
\ee
Solving (\ref{massshellflow}) for $m$, and plugging into (\ref{3.2}) we obtain 
\be\label{hNS}
h = \frac{N}{w} + \frac{h_{\rm rest}  + s^2}{w} +  \frac{w^2-1 }{4 w}\ . 
\ee
This quantity is to be identified with the (left-moving) conformal dimension $h$ in the dual CFT. There is also a similar formula for the right-moving conformal dimension
\be\label{barh}
\bar{h} = \bar{m} + \frac{w}{2} =\frac{\bar{N}}{w} + \frac{\bar{h}_{\rm rest}  + s^2}{w} + \frac{w^2-1}{4w} \ . 
\ee
Note that since the left- and right-moving representation must be the `same' continuous representation, it follows that $m-\bar{m} \in \mathbb{Z}$. As a consequence, the left- and right-moving conformal dimensions of the dual CFT differ by an integer. 

In the R sector the analysis is essentially identical --- recall, in particular, that $m$ refers to the $J^3_0$ eigenvalue with respect to the original (coupled) generator --- the only difference being that the right--hand-side of (\ref{massshellflow}) is $0$ rather than $\frac{1}{2}$. Thus we get from the R sector the identities 
\begin{eqnarray}
h & =  & m + \frac{w}{2} = \frac{N}{w} + \frac{h_{\rm rest}  + s^2}{w}  +  \frac{w^2+1}{4w}\ \\[2pt]
\bar{h} & = & \bar{m} + \frac{w}{2} = \frac{\bar{N}}{w} + \frac{h_{\rm rest}  + s^2}{w} + \frac{w^2+1}{4w} \ .
\end{eqnarray}

\subsection{A Special Subsector}

The spectrum becomes particularly simple if we consider the states with $h_{\rm rest}=0$ for the lowest continuous representation, i.e., the representation with $s=0$. We should mention that the restriction to $h_{\rm rest}=0$ means, in particular, that the winding and momentum modes along the circle are set to zero. As a consequence, the dual CFT will also not have any momentum for the free boson. It would be interesting to generalise our analysis to include also these modes, but we have not attempted this so far. 

The following analysis depends a bit on whether $w$ is even or odd; we shall therefore discuss the two cases in turn.

\subsubsection{Odd Spectral Flow }

For odd spectral flow, we need to impose the GSO projection that $N$ has to be an integer in the NS sector,\footnote{In the R sector, the sign of the GSO projection does not affect the counting function because of the zero modes.} see eq.~(\ref{GSOw}). Recalling that we have $8$ free fermions and $2$ free bosons, it follows from eq.~(\ref{hNS})  that the generating function of the (left-moving) physical states is then of the form 
\begin{eqnarray}
Z & = & {\displaystyle q^{h_w}\, \prod_{n=1}^\infty \frac{1}{(1-q^{\frac{n}{w}})^2} \, \frac{1}{2}
\Bigl[ \prod_{n=1}^{\infty} (1 + q^{\frac{2n-1}{2w}})^8 + \prod_{n=1}^{\infty} (1 - q^{\frac{2n-1}{2w}})^8 + 16 \,q^{\frac{1}{2w}} \prod_{n=1}^{\infty} 
(1+q^{\frac{n}{w}})^8 \Bigr] } 
\nonumber \\[4pt] 
& = & {\displaystyle q^{h_w} \, \prod_{n=1}^\infty \frac{1}{(1-q^{\frac{n}{w}})^2} \,  \prod_{n=1}^{\infty} (1 + q^{\frac{2n-1}{2w}})^8  \ ,} \label{NSmain}
\end{eqnarray}
where
\be\label{oddwground}
h_w = \frac{w^2-1}{4w}  \qquad \hbox{($w$ odd)}\ ,
\ee
and the coefficient of $q^n$ is the number of physical states with $h=n$. In going to the second line of (\ref{NSmain}) we have used  the so-called abstruse identity of Jacobi theta functions, see e.g.\ \cite{WW}. We note that the second line of (\ref{NSmain}) looks like the partition function of eight NS-sector fermions and two bosons with fractional modes, but without any GSO-projection. We will be more specific about this below in Section~\ref{sec:symorbodd}.

It is also useful to keep track of the chemical potentials of the $\mathfrak{su}(2)$ factors. In particular, we can replace the bracket in the first line of eq.~(\ref{NSmain}) by 
\begin{eqnarray}
& & \frac{1}{2}\, \Bigl[ \prod_{n=1}^{\infty} (1 +  y\, q^{\frac{2n-1}{2 w}}) \, (1 +  y^{-1} q^{\frac{2n-1}{2 w}})\, (1 +  z \, q^{\frac{2n-1}{2 w}}) \, (1 +  z^{-1} q^{\frac{2n-1}{2 w}})   \, (1 +  q^{\frac{2n-1}{2 w}})^4  \nonumber \\
& & \quad + \prod_{n=1}^{\infty} (1 -  y \, q^{\frac{2n-1}{2 w}}) \, (1 -  y^{-1} q^{\frac{2n-1}{2 w}})\, (1 -  z \, q^{\frac{2n-1}{2 w}}) \, (1 -  z^{-1} q^{\frac{2n-1}{2 w}})   \, (1 -  q^{\frac{2n-1}{2 w}})^4 \qquad\quad  \nonumber % \\
\end{eqnarray}
\begin{eqnarray}
& & \quad + q^{\frac{1}{2w}}\, 4 \, (y^{\frac{1}{2}}+ y^{-\frac{1}{2}})\, (z^{\frac{1}{2}}+ z^{-\frac{1}{2}}) \, \times \nonumber \\  
& &\qquad \qquad  \times \prod_{n=1}^{\infty} (1+y \, q^{\frac{n}{w}}) \, (1+y^{-1}q^{\frac{n}{w}}) \, (1+z \, q^{\frac{n}{w}}) \, (1+z^{-1}q^{\frac{n}{w}}) \, 
(1+q^{\frac{n}{w}})^4 \Bigr] \label{abstrusecharge}  \\
& & = \prod_{n=1}^{\infty} (1 + y^{\frac{1}{2}}  z^{\frac{1}{2}}  q^{\frac{2n-1}{2w}})^2\, (1 + y^{\frac{1}{2}}  z^{-\frac{1}{2}}  q^{\frac{2n-1}{2w}})^2\,
(1 + y^{-\frac{1}{2}} z^{\frac{1}{2}}  q^{\frac{2n-1}{2w}})^2\, (1 + y^{-\frac{1}{2}} z^{-\frac{1}{2}}  q^{\frac{2n-1}{2w}})^2 \ . \nonumber 
\end{eqnarray}
Here we have introduced a chemical potential for each of the two $\mathfrak{su}(2)$ factors, corresponding to the two ${\rm S}^3$ factors. (Recall that for each $\mathfrak{su}(2)$ algebra, the free fermions $\chi^a$ transform in the adjoint representation, and hence carry charge $0,\pm 1$.)
This identity can be derived using Jacobi's addition formulas for theta functions (see e.g.\  page 487 of \cite{WW}). Note that (\ref{abstrusecharge}) reduces to the abstruse identity that we used in going to the second line of (\ref{NSmain}) for $y=z=1$.

\subsubsection{Even Spectral Flow } 

The analysis for even $w$ is similar, the only difference being that now $N$ has to be half-integer in the NS sector, see eq.~(\ref{GSOw}). Thus instead of (\ref{NSmain}), we get 
\begin{eqnarray}
Z & = & {\displaystyle q^{h_w}\, \prod_{n=1}^\infty \frac{1}{(1-q^{\frac{n}{w}})^2} \, \frac{1}{2}
\Bigl[ \prod_{n=1}^{\infty} (1 + q^{\frac{2n-1}{2w}})^8 - \prod_{n=1}^{\infty} (1 - q^{\frac{2n-1}{2w}})^8 + 16 \,q^{\frac{1}{2w}} \prod_{n=1}^{\infty} 
(1+q^{\frac{n}{w}})^8 \Bigr] } 
\nonumber \\[4pt] 
& = & {\displaystyle 16\, q^{h_w'}\,  \prod_{n=1}^\infty \frac{1}{(1-q^{\frac{n}{w}})^2} \,  \prod_{n=1}^{\infty} (1 + q^{\frac{n}{w}})^8 \ ,} \label{Rmain}
\end{eqnarray}
where
\be\label{evenwground}
h_w' = \frac{w^2-1}{4w} + \frac{1}{2w} = \frac{w^2+1}{4w} \qquad \hbox{($w$ even)}\ . 
\ee
This now looks like the partition function of eight R-sector fermions and two bosons with fractional modes, but without any GSO-projection; again, we will come back to the more detailed interpretation below in Section~\ref{sec:symorbeven}. The relevant generalisation involving chemical potentials is now 
\begin{eqnarray}
& & \frac{1}{2}\, \Bigl[ \prod_{n=1}^{\infty} (1 +  y \, q^{\frac{2n-1}{2 w}}) \, (1 +  y^{-1} q^{\frac{2n-1}{2 w}})\, (1 +  z \, q^{\frac{2n-1}{2 w}}) \, (1 +  z^{-1} q^{\frac{2n-1}{2 w}})   \, (1 +  q^{\frac{2n-1}{2 w}})^4  \nonumber \\
& & \quad - \prod_{n=1}^{\infty} (1 -  y \, q^{\frac{2n-1}{2 w}}) \, (1 -  y^{-1} q^{\frac{2n-1}{2 w}})\, (1 -  z \, q^{\frac{2n-1}{2 w}}) \, (1 -  z^{-1} q^{\frac{2n-1}{2 w}})   \, (1 +  q^{\frac{2n-1}{2 w}})^4  \nonumber \\
& & \quad + q^{\frac{1}{2w}}\, 4 \, (y^{\frac{1}{2}}+ y^{-\frac{1}{2}})\, (z^{\frac{1}{2}}+ z^{-\frac{1}{2}}) \times \nonumber \\ 
& & \qquad \qquad \qquad \times\prod_{n=1}^{\infty} (1+y \, q^{\frac{n}{w}}) \, (1+y^{-1}q^{\frac{n}{w}}) \, (1+z \, q^{\frac{n}{w}}) \, (1+z^{-1}q^{\frac{n}{w}}) \, 
(1+q^{\frac{n}{w}})^4 \Bigr] \nonumber \\
& & = q^{\frac{1}{2w}}\, \bigl(y^{\frac{1}{2}} z^{\frac{1}{2}}+ y^{-\frac{1}{2}}z^{-\frac{1}{2}}\bigr)^2 \,   
\bigl(y^{\frac{1}{2}} z^{-\frac{1}{2}}+ y^{-\frac{1}{2}}z^{\frac{1}{2}}\bigr)^2\times   \label{abstrusecharge2} \\ 
& & \quad  \times \prod_{n=1}^{\infty} (1+y^{\frac{1}{2}}z^{\frac{1}{2}} \, q^{\frac{n}{w}})^2 \, (1+y^{\frac{1}{2}}z^{-\frac{1}{2}}q^{\frac{n}{w}})^2 \, (1+y^{-\frac{1}{2}}z^{\frac{1}{2}} \, q^{\frac{n}{w}})^2 \, (1+y^{-\frac{1}{2}}z^{-\frac{1}{2}}q^{\frac{n}{w}})^2 \ , \nonumber
\end{eqnarray}
as follows by similar arguments to (\ref{abstrusecharge}), see \cite{WW}.

\section{Matching with the Symmetric Orbifold Spectrum}\label{sec:symorb}

In this section we want to show that these generating functions reproduce precisely the single-particle spectrum of the symmetric orbifold of $8$ free fermions and $2$ free bosons, i.e., of the $({\cal S}_0)^2$ theory, in the (spacetime) NS sector. (The R sector states of the dual CFT are expected to correspond to non-perturbative excitations, such as black holes, and will not be directly visible from the world-sheet perspective.)  We shall first consider the untwisted sector states, and then study the different twisted sectors. 

\subsection{The Single-Particle States from the Untwisted Sector}

We begin with explaining the structure of the single-particle states that arise from the untwisted sector of the symmetric orbifold of the seed theory ${\cal H}_0$. (In our case, the seed theory will be the NS sector of $8$ free fermions and $2$ free bosons.)  The untwisted sector consists of all $S_N$ invariant states of $({\cal H}_0)^N$. Not all of these states should be thought of as `single-particle' states though. Indeed, as is for example explained in \cite[Section 2]{Gaberdiel:2015mra}, the full contribution of the untwisted sector is, for large $N$, naturally in `multi-particle' form, and the single-particle states are counted precisely by the partition function of ${\cal H}_0$ itself. 

For the case at hand, the single-particle states from the untwisted sector of the symmetric orbifold are therefore counted by the partition function 
\be
Z_0 = \prod_{n=1}^\infty \frac{1}{(1-q^{n})^2} \,  \prod_{n=1}^{\infty} (1 + q^{\frac{2n-1}{2}})^8 \ . 
\ee
This agrees indeed precisely with the partition function (\ref{NSmain}) for $w=1$. Thus the sector with $w=1$ corresponds to the untwisted sector of the symmetric orbifold.

\subsection{Sectors of Odd Cycle Length}\label{sec:symorbodd}

The other single-particle states of the symmetric orbifold arise from single-cycle twisted sectors. (Twisted sectors associated to permutations with more than one cycle are multi-particle.) The analysis depends a bit on whether the cycle length $L$ is even or odd; we shall first deal with the slightly simpler case where $L$ is odd. As explained in Appendix~\ref{app:ground}, the ground state energy of the symmetric orbifold of $8$ free fermions and $2$ free bosons then takes the form 
\be
h_0 = \frac{L^2-1}{4L} \ .
\ee
Furthermore, it is clear from the derivation there that the single-particle descendants of this ground state are counted by the partition function of ${\cal H}_0$ evaluated at $q^{\frac{1}{L}}$.\footnote{There are also multi-particle states coming from multiplying these descendants with invariant states from the copies unaffected by the twist.} The relevant single-particle contribution is therefore for odd cycle length $L$
\be
Z_L = q^{\frac{L^2-1}{4L}} \, \prod_{n=1}^\infty \frac{1}{(1-q^{\frac{n}{L}})^2} \,  \prod_{n=1}^{\infty} (1 + q^{\frac{2n-1}{2L}})^8 \ . 
\ee
This agrees precisely the partition function (\ref{NSmain}) for $w=L$ odd. 

\subsection{Sectors of Even Cycle Length}\label{sec:symorbeven}

The analysis for even cycle length is slightly more subtle since the partition function of the twisted sector then looks like a Ramond sector, see the discussion below eq.~(\ref{twinedsectorp}). Furthermore, the ground state energy for the case of $8$ free fermions and $2$ free bosons equals (see eq.~(\ref{Levenground})) 
\be\label{Levengroundm}
h_0 = \frac{L^2+1}{4L} \qquad \hbox{($L$ even)}  \ .
\ee
Thus the single-particle contribution is for even cycle length $L$
\be
Z_L = 16\, q^{\frac{L^2+1}{4L}} \, \prod_{n=1}^\infty \frac{1}{(1-q^{\frac{n}{L}})^2} \,  \prod_{n=1}^{\infty} (1 + q^{\frac{n}{L}})^8 \ ,
\ee
where the prefactor of $16$ comes from the fact that there are $8$ fermionic zero modes, leading to a $16$ dimensional Clifford representation. This then 
agrees precisely with the partition function (\ref{Rmain}) for $w=L$ even.

\subsection{Orbifold Projection and Chemical Potentials}

So far we have only considered one chiral half of these theories, but in the full theory we also have to impose the orbifold projection. Thus we need to combine the left- and right-moving states so that they are invariant under the centraliser of the twist. This is equivalent to demanding that $h-\bar{h}\in\mathbb{Z}$, which is also true in the world-sheet description, see the discussion below eq.~(\ref{barh}). Note that this is the correct prescription in the NS-NS sector of the world-sheet that corresponds to bosonic degrees of freedom of the dual CFT. Because of the shift in $J^3_0$ in going from the NS-sector to the R-sector, see e.g., eq.~(2.17) in \cite{Ferreira:2017pgt}, the relevant condition in the NS-R sector (that corresponds to fermions in the dual CFT) is $h-\bar{h}\in\mathbb{Z}+\frac{1}{2}$, and similarly for the other sectors.

We can also keep track of the chemical potential for the two $\mathfrak{su}(2)$ algebras, see the discussion around eqs.~(\ref{abstrusecharge}) and (\ref{abstrusecharge2}). From the perspective of the dual CFT, the eight free fermions then transform as $2 \cdot ({\bf 2},{\bf 2})$ with respect to these two $\mathfrak{su}(2)$ algebras, see eqs.~(\ref{abstrusecharge}) and (\ref{abstrusecharge2}). Thus the seed theory of the symmetric orbifold has exactly the structure of $({\cal S}_0)^2$, where ${\cal S}_0$ is the theory of one real boson and $4$ real fermions, transforming as $({\bf 2},{\bf 2})$ with respect to the two $\mathfrak{su}(2)$ algebras of the large ${\cal N}=4$ superconformal algebra, see, e.g., \cite{Gukov:2004ym}. The spacetime spectrum of the $k=1$ theory is therefore very similar to the symmetric orbifold of ${\cal S}_0$ that was proposed to be the CFT dual of this background (for the case where $k_+=k_-$) in \cite{Eberhardt:2017pty}.

\subsection{Further Comments}

One may wonder whether our findings suggest that the correct CFT dual of this background is the symmetric orbifold of $({\cal S}_0)^2$, rather than that of ${\cal S}_0$, as proposed in \cite{Eberhardt:2017pty}. However, we suspect that this is not the correct conclusion, for the following reason. The BPS spectrum of the $k=1$ world-sheet theory has far fewer BPS states than, say, the supergravity spectum of \cite{Eberhardt:2017fsi}, and this is mirrored by the symmetric orbifold of $({\cal S}_0)^2$. 
 Indeed, as is also familiar for the case of $\mathbb{T}^4$, the BPS spectrum of the WZW world-sheet theory has gaps, and for $k=1$ they are quite frequent and remove in fact  all BPS states with $j^+=j^- =$ half-integer, see the discussion in appendix~E of \cite{Eberhardt:2017pty}. 
 
In particular, neither the $k=1$ world-sheet theory (nor the symmetric orbifold of $({\cal S}_0)^2$), has an exactly marginal operator since the BPS state with $j^+=j^-=\frac{1}{2}$ is missing.  This should be contrasted with the symmetric orbifold of ${\cal S}_0$ that does possess such an operator. Indeed, as explained in  \cite{Eberhardt:2017pty}, see also \cite{Gukov:2004ym}, this marginal operator arises from the $3$-cycle twist ---  there are no BPS states in the symmetric orbifold of ${\cal S}_0$ for even cycle length. (The reason why there are no BPS states for even cycle length is exactly the same as we saw in our analysis above, namely that for even twist the ground state energy, see eq.~(\ref{Levengroundm}), is too high.) Note that for our symmetric orbifold of $8$ free fermions and $2$ free bosons (i.e.\ for $({\cal S}_0)^2$), while there is a BPS state in the 3-cycle twisted sector, it does not correspond to an exactly marginal operator since, from the perspective of a single ${\cal S}_0$ theory, this is like the twisted sector corresponding to 2 $3$-cycles. And while this is BPS, it has simply double the conformal weight und $\mathfrak{su}(2)$ charges. 

On the other hand, the symmetric orbifold of $({\cal S}_0)^2$ does look as though it could be the IR fixed point of the brane construction considered in \cite{Tong:2014yna}.\footnote{We thank David Tong for a useful discussion about this point.} The fact that it does not have an exactly marginal operator in its spectrum seems to indicate that it is an infinite distance away in moduli space. This is probably also related to the  instanton singularity of \cite{Seiberg:1999xz}.

\section{The Case of $\mathbb{T}^4$}\label{sec:symorbT4}

For the case of ${\rm AdS}_3\times {\rm S}^3 \times \mathbb{T}^4$ the situation is slightly more subtle. As was reviewed in Section~\ref{sec:worldsheet}, for $k=1$ the decoupled bosonic $\mathfrak{su}(2)$ algebra is at level $\kappa=-1$. On the face of it, this leads to an inconsistent world-sheet theory, and for this reason, the case $k=1$ has often been discarded. In the following we want to argue that there is a fairly natural way in which one can make sense of this theory by observing that $\mathfrak{su}(2)$  at level $\kappa=-1$ has a free field construction in terms of four symplectic bosons \cite{Goddard:1987td}, see also \cite[eq.~(3.19)]{Beem:2014rza},  that effectively behave as fermionic ghosts.  Since the symplectic boson construction is not very well known, we shall briefly review it below. 

\subsection{The Complex Fermion Construction}

The symplectic boson construction is the natural analogue of the more familiar free field realisation of $\mathfrak{su}(2)_1$ in terms of two complex fermions, $\chi^i$ and $\bar{\chi}^i$, $i=1,2$, satisfying 
\be\label{ffanti}
\{\chi^i_r,\bar{\chi}^j_s\} = \delta^{ij} \, \delta_{r,-s} \qquad \{\chi^i_r,\chi^j_s\} = \{\bar{\chi}^i_r,\bar{\chi}^j_s \} = 0 \ .
\ee
In this case, one defines the $\mathfrak{su}(2)$ currents as 
\be
J^a = t^a_{ij} \, \chi^i \bar{\chi}^j \ , 
\ee
where $t^a_{ij}$ are the representation matrices of $\mathfrak{su}(2)$ in the ${\bf 2}$-dimensional $j=\frac{1}{2}$ representation. These generators give rise to $\mathfrak{su}(2)$ at level $k=1$. In addition, there is a decoupled $\mathfrak{u}(1)$ current corresponding to 
\be
U = \chi^i \, \bar{\chi}^i \ ,
\ee
reflecting the fact that the complex fermions naturally lead to $\mathfrak{u}(2) \cong \mathfrak{su}(2) \oplus \mathfrak{u}(1)$. For the case at hand, the 
$\mathfrak{u}(1)$ current $U$ can actually be extended to another $\mathfrak{su}(2)_1$ algebra --- we can equivalently think of this construction in terms of $4$ real fermions generating $\mathfrak{so}(4) \cong \mathfrak{su}(2) \oplus \mathfrak{su}(2)$ --- by considering the charged generators
\be
K^+ = \chi^1 \chi^2 \ , \qquad K^- = \bar{\chi}^1 \bar{\chi}^2 \ . 
\ee
This then accounts for the full central charge of $c=2$ coming from $2$ complex or $4$ real fermions. The representations of $\mathfrak{su}(2)_1\oplus \mathfrak{su}(2)_1$ are also naturally described in this language: in the NS sector we have 
\be\label{NSf}
{\cal H}_{{\rm NS}} = \bigl( {\cal H}_{j=0} \otimes {\cal H}_{j=0}   \bigr) \oplus \bigl( {\cal H}_{j=\frac{1}{2}} \otimes  {\cal H}_{j=\frac{1}{2}}  \bigr) \ ,
\ee
while the R sector leads to 
\be\label{Rf}
{\cal H}_{{\rm R}}  =  \bigl( {\cal H}_{j=\frac{1}{2}} \otimes {\cal H}_{j=0}   \bigr) \oplus  \bigl( {\cal H}_{j=0} \otimes {\cal H}_{j=\frac{1}{2}}   \bigr)  \ .
\ee
Note, in particular, that the $j=\frac{1}{2}$ representation of $\mathfrak{su}(2)_1$ has conformal dimension $h=\frac{1}{4}$, which fits with these assignments. (For example, the conformal dimension of the ground state of the second summand in (\ref{NSf}) is then $h=\frac{1}{2} = \frac{1}{4} + \frac{1}{4}$, while that of either term in (\ref{Rf}) is $h=\frac{1}{4}+0 = \frac{1}{4}$.)

\subsection{The Symplectic Boson Construction}

For the symplectic boson we proceed similarly, except that instead of free fermions we now have symplectic boson fields $\xi^{i}$ and $\bar{\xi}^j$ $(i,j=1,2)$, satisfying  \cite{Goddard:1987td}
\be
[\xi^i_r,\bar{\xi}^j_s] = \delta^{ij}\, \delta_{r,-s} \ , \qquad [\xi^i_r,\xi^j_s] = [\bar{\xi}^i_r,\bar{\xi}^j_s] =  0 \ ,
\ee
i.e., instead of the anti-commutators in (\ref{ffanti}) we now have commutation relations. Thus the underlying fields are bosons, but they have spin $s=\frac{1}{2}$, as one can read off from the $(r,s)$ dependence of these relations --- indeed, it is the same as for the fermions above. The currents can now be defined as
\be
J^a = - t^a_{ij} \, \xi^i \bar{\xi}^j \ , 
\ee
and they lead to $\mathfrak{su}(2)$ at level $\kappa=-1$, as one can check explicitly, see also \cite[eq.~(3.19)]{Beem:2014rza}. This fits with the general relation, see eq.~(2.14) of  \cite{Goddard:1987td}, \be\label{kappap}
- \kappa  \dim(\mathfrak{su}(2)) =   Q_R \, \dim(R) = \frac{3}{4}\, 4  = 3 \ ,
\ee
where $Q_R$ is the value of the quadratic Casimir in the representation $R$. In our case, $R$ is $4$-dimensional, consisting of two copies of the $j=\frac{1}{2}$ representation, and hence $Q_R = j (j+1) = \frac{3}{4}$. Furthermore, while there still exists a decoupled $\mathfrak{u}(1)$ current associated to 
\be
U = \xi^i \bar{\xi}^i \ , 
\ee
it is now not possible to extend this to a commuting $\mathfrak{su}(2)$ algebra since the analogues  of $K^\pm$ do not commute with the generators $J^\pm$. (While, for example, $\chi^1\chi^1=0$ by virtue of the anti-commutation relations, the same is not true for $\xi^1\xi^1$.) Thus the relevant algebra is 
\be\label{su2m1}
\mathfrak{su}(2)_{-1} \oplus \mathfrak{u}(1) \ , 
\ee
and this has $c=-3+1 = -2$, in agreement with the central charge of the four symplectic bosons. (Each symplectic boson constributes $c=-\frac{1}{2}$.)

There are again two natural sectors: an NS-like sector in which the symplectic bosons are half-integer moded (and that contains the vacuum representation of $\mathfrak{su}(2)_{-1}$), as well as an R-like sector in which the symplectic bosons are integer moded. It would be interesting to study the analogues of (\ref{NSf}) and (\ref{Rf}) in terms of the $\mathfrak{su}(2)_{-1}\oplus \mathfrak{u}(1)$ representation theory, but we have not done so yet since this is not directly needed for our purposes here. The only fact that will be relevant below is that the ground state of the R-sector representation has conformal dimension $h=-\frac{1}{4}$. Indeed, as explained  below eq.~(1.30) of \cite{Goddard:1987td}, for each symplectic boson the R-sector energy is lower by $\delta h = -\frac{1}{16}$, and since we have four symplectic bosons we arrive in total at $h=-\frac{1}{4}$.

\subsection{The World-sheet Theory}

After these preparations we can now return to the problem at hand, namely to make sense of the world-sheet theory of ${\rm AdS}_3\times {\rm S}^3 \times \mathbb{T}^4$ at $k=1$. The basic idea is that we can represent the problematic $\mathfrak{su}(2)_{-1}$ factor in terms of four symplectic bosons. In fact, since the symplectic bosons generate in addition one $\mathfrak{u}(1)$ factor, see eq.~(\ref{su2m1}), the resulting degrees of freedom of our world-sheet theory are then $6$ bosons ($3$ from $\mathfrak{sl}(2,\mathbb{R})$, as well as $4-1$ from the torus), together with the $4$ symplectic bosons from above. The fermions are unmodified, and we continue to have $10$ fermionic degrees of freedom.  After the physical state condition is imposed this then leads to $4$ bosonic and $8$ fermionic degrees of freedom (as well as the $4$ symplectic bosons). 

The bosonic degrees of freedom are unaffected by the world-sheet GSO projection, and hence go along for the ride in the manipulations of eqs.~(\ref{NSmain}) or (\ref{Rmain}). (Indeed, since this world-sheet theory has the same fermionic degrees of freedom as for the case of ${\rm AdS}_3 \times {\rm S}^3 \times {\rm S}^3 \times {\rm S}^1$, these manipulations apply as before.) In particular, the resulting spacetime spectrum has then the form of a symmetric orbifold of $8$ free fermions, $4$ free bosons, as well as the $4$ symplectic bosons. 

Next we note that the four symplectic bosons behave effectively as fermionic ghosts  \cite{Goddard:1987td}. Thus the effect of these four symplectic bosons is to remove $4$ of the free fermions,\footnote{We thank Lorenz Eberhardt for discussions about a related idea.} so that we end up with $4$ free bosons and fermions, i.e.\ the $\mathbb{T}^4$ theory. 
We should note that this also fits with the charges under the $\mathfrak{su}(2)$ algebra corresponding to the ${\rm S}^3$ factor. 
It follows from (\ref{abstrusecharge}) and (\ref{abstrusecharge2}) for $z=1$ --- since there is no second ${\rm S}^3$ factor, we should set the corresponding chemical potential to $z=1$ --- that in the dual CFT the eight fermions consist of $4$ doublets with respect to that $\mathfrak{su}(2)$. This ties together with the fact that the four symplectic bosons also sit in $2$ doublets with respect to that $\mathfrak{su}(2)$, see the comment below (\ref{kappap}).

There is however one important subtlety. It follows from eqs.~(\ref{NSmain}) and (\ref{Rmain}) that the free fermions are in the NS-like sector for $w$ odd, and in the R-like sector for $w$ even. In order for the symplectic bosons to cancel against $4$ of the $8$ fermions, they must behave in the same manner. Thus we are led to propose that the spectrum of our world-sheet theory consists of $4$ symplectic bosons that are half-integer moded for $w$ odd, and integer-moded for $w$ even.\footnote{More fundamentally, this selection rule should follow from the requirement of the theory to respect spacetime supersymmetry, and it would be very interesting to understand it from this perspective.} We should mention that choosing different representations for $w$ even and $w$ odd is maybe not too unnatural. In particular,  for the analysis of the BPS spectrum for the $\mathbb{T}^4$ world-sheet theory (for generic $k$), see in particular \cite{Argurio:2000tb,Raju:2007uj}, it is convenient to also spectrally flow in the $\mathfrak{su}(2)$ sector. This is (for positive integer $k$) an involution, i.e., it only depends on the spectral flow sector mod $2$.  Thus also from this perspective even and odd $w$ behave slightly differently.

The above prescription for the moding of the symplectic bosons has also another desirable consequence. For $w$ even, the contribution of $h_{\rm rest}$ is then 
$h=-\frac{1}{4}$ (see the discussion at the end of the previous subsection) and as a consequence the ground state energy in (\ref{evenwground}) is modified to 
\be\label{groundevenT4}
\hbox{$w$ even:} \qquad \frac{w^2- 1}{4w} + \frac{1}{2w} - \frac{1}{4w}  = \frac{w}{4} \ ,
\ee
in agreement with (\ref{LevenT4}) for $L=w$. (Without this contribution of $h_{\rm rest}=-\frac{1}{4}$, the ground state energy would have been again (\ref{Levengroundm}).)
 Thus, with this prescription for the world-sheet theory, we reproduce exactly the symmetric orbifold spectrum of four free bosons and fermions.

\section{Discussion}\label{sec:concl}

In trying to understand the different possible tensionless limits of string theory on AdS$_3$, we have found an unexpected congruence between two apparently different points in parameter space: the spectrum of the theory with NS-NS flux, at $k=1$, contains a natural subsector which is identical to that of the symmetric product orbifold.\footnote{At first sight the observation here seems to be similar to one made in \cite{Argurio:2000tb}. However, there are some key differences. The proposal in \cite{Argurio:2000tb} is for generic $k$, while ours is restricted to the special value $k=1$. Their comparison is largely about the chiral primary states  coming from the discrete representations (short strings), while our result concerns the full symmetric orbifold spectrum and we are focussing on the subsector sitting at the bottom of the continuous representations. It will be interesting, nevertheless, to understand any connection between the two proposals.} This observation raises several questions in turn, the first being -- why? A plausible answer, mentioned in the introduction, is that the spectrum is largely dictated by the HSS enhanced symmetry. As was discussed in \cite{Gaberdiel:2015wpo} (see also \cite{Raeymaekers:2016mmm, Sharma:2017zjf}), the structure of the untwisted sector of the symmetric orbifold is very much like that of the Vasiliev higher spin theory in having a higher spin tower coupled to a single massive minimal representation. In addition, the twisted sectors are also natural  near-minimal representations. It is conceivable that one needs to have this particular set of representations in order to have a consistent theory. This would suggest that this structure is rigid and must be present at other tensionless points as well. 

If this is indeed the correct answer, we can try to test this in a number of ways. We could check that this subsector is closed (in the sense of OPEs) from the point of view of the world-sheet theory. Note that the correspondence with the twisted sectors implies specific fusion rules between the different spectrally flowed continuous representations. 
One may also try to look for the HSS operating on the level of the world-sheet. This may shed some light also on how this symmetry becomes unbroken (unhiggsed) at the special point $k=1$. 

Another interesting line of thought is that there is something topological about this subsector (or at least the massless HSS generators), and that we can follow these states even after deforming away from the pure NS-NS point. (Indeed, the ground states of the continuous representations seem to correspond to some sort of world-sheet instantons since they correspond to holomorphic maps from the world-sheet torus to the boundary torus (at finite temperature), see the discussion around eq.~(73) of \cite{Maldacena:2000kv}.\footnote{See also \cite{Rastelli:2005ph} for an identification of a somewhat different topological sector in the AdS$_3$ string theory.})
This would be very exciting if true, and enable one to try to extrapolate in some sense from this point to the orbifold point. In this context, we should note that the pure NS-NS background and the symmetric orbifold behave somewhat asymmetrically. In particular, there are no deformations at the symmetric orbifold point which correspond to turning on NS-NS flux --- after all NS-NS flux is quantised --- while one can deform the NS-NS world-sheet  theory with a R-R flux deformation because the relevant parameter is $\lambda Q_{\rm RR}$ (where $\lambda$ is the ten-dimensional string coupling constant) which one may take to be continuous, see the discussion in \cite{Berkovits:1999im}.\footnote{We thank Nick Dorey for discussions about this point.}

Finally, assuming that the rest of the continuum of the continuous representations are lifted once one goes away from the pure NS-NS flux case, there is still a question about the role of the other physical states coming from spectrally flowed discrete representations, as mentioned in Section~\ref{sec:2.2}. Given that we know that the states from the symmetric product orbifold gives rise to a modular invariant spacetime spectrum, it seems a little odd that there is freedom to add in some additional states. In order to understand their significance it may also be instructive to understand how they fit into representations of the HSS.
\smallskip

As a side product of our analysis we have also made a proposal for how to make sense of the world-sheet theory for ${\rm AdS}_3\times {\rm S}^3 \times \mathbb{T}^4$ at $k=1$, where the decoupled bosonic $\mathfrak{su}(2)$ algebra appears at level $\kappa=-1$. It would be interesting to scrutinize this further; in particular, it would be interesting to describe the symplectic boson representations in terms of the $\mathfrak{su}(2)_{-1}$ representation theory --- since $\kappa=-1$ is not admissible in the sense of \cite{KW}, there is no natural class of representations that one may expect to appear --- and check that the resulting world-sheet theory is indeed modular invariant. 
Furthermore, it would be interesting to confirm that the specific selection rules on the spectrum, see the discussion before eq.~(\ref{groundevenT4}), are a consequence of requiring spacetime supersymmetry.
\bigskip

\noindent {\bf Note added.} A preliminary version of this work was presented by RG at \cite{RG}. It was subsequently brought to our attention that related work has independently been done in \cite{GHKPR}. We thank the authors of  \cite{GHKPR} for sharing their draft with us prior to publication.

\section*{Acknowledgements}

We thank Nick Dorey, Lorenz Eberhardt, David Kutasov, Wei Li, Juan Maldacena, Massimo Porrati, Leonardo Rastelli, Alessandro Sfondrini and David Tong for useful discussions. MRG thanks the BICMR at Beijing University for hospitality while the bulk of this work was done. He was also supported in part by the NCCR SwissMAP, funded by the Swiss National Science Foundation. RG's research was supported in part by the J.C.~Bose Fellowship of the SERB, Govt.~of India, and the Infosys Excellence Grant to the ICTS.  He would like to specially acknowledge the overarching framework of support for the basic sciences from the people of India.  

\appendix

\section{The Zero Point Energy of the $L$-cycle Sector}\label{app:ground}

Let us begin by considering the symmetric orbifold of a single free boson. The chiral part of the partition function then goes as 
\begin{equation}\label{onebos}
{\rm Tr}_{{\cal H}^{(1)}} \bigl( q^{L_0 - \frac{1}{24}} \bigr) =
q^{-\frac{1}{24}} \bigl( 1 + \cdots \bigr) = \chi(\tau) \ ,
\end{equation}
where $q=e^{2\pi i \tau}$. (For the case of the free boson we obviously just have $\chi(\tau) = \eta(\tau)$, but this will actually not be very significant in the following.) 

Now consider the $N$-fold product theory, whose
space of states we denote by ${\cal H}^{(N)}$. Let $\sigma$ be the cyclic permutation of length $L$, 
then the twining character, i.e., the character over ${\cal H}^{(N)}$ with the insertion of $\sigma$,
equals
\begin{equation}\label{twinedsector}
{\rm Tr}_{{\cal H}^{(N)}} \Bigl(\sigma q^{L_0 - \frac{N}{24}} \Bigr) = \chi(L\tau)\, \chi(\tau)^{N-L} \ . 
\end{equation}
(This is just a consequence of the fact that only the totally symmetric states of the first $L$ copies contribute to
this twining character.)

In order to obtain the character in the twisted sector, we now need to take the S-modular transformation, 
which therefore goes as 
\begin{equation}
Z_T = \chi \Bigl(\frac{\tau}{L}\Bigr)\, \chi(\tau)^{N-L}  = q^{-\frac{1}{24 L}} q^{-\frac{N-L}{24}}\, \bigl( 1 + \cdots \bigr) \ .
\end{equation}
The leading exponent equals the zero point energy in the twisted sector minus $\tfrac{N}{24}$, i.e., 
we have
\begin{equation}\label{zerob}
h_0 =  \frac{L}{24} - \frac{1}{24 L}  = \frac{L^2-1}{24 L} \ . 
\end{equation}
\smallskip

The analysis for the fermions is essentially identical, except that one has to be careful about whether
$L$ is even or odd. In the simpler case, i.e., when $L$ is odd, the only difference to the above is that the 
fermionic analogue of (\ref{onebos}) goes as $q^{-1/48}$ in the NS sector. Thus everything for fermions
is precisely halved. Thus for a theory consisting of $B$ free bosons and $F$ free fermions, the ground state energy in the  $L$-cycle twisted sector is given by 
\be\label{h0Lodd}
h_0 = (2 B + F) \,  \frac{L^2-1}{48 L}  \qquad \hbox{$L$ odd} \ . 
\ee
For example, for the case of $\mathbb{T}^4$ for which $B=F=4$, this leads to $h_0 = \frac{1}{4L} (L^2-1)$, in agreement, for example, 
with eq.\ (2.11) in   \cite{Lunin:2001pw}. Incidentally, we also find the same result for $F=8$ and $B=2$, 
\be\label{Loddground}
h_0 = \frac{L^2-1}{4L} \qquad \hbox{($L$ odd)}  \ .
\ee
\medskip

\noindent For even $L$ there is a subtlety in that the analogue of (\ref{twinedsector}) is 
\begin{equation}\label{twinedsectorp}
{\rm Tr}_{{\cal H}^{(N)}} \Bigl(\sigma q^{L_0 - \frac{N}{24}} \Bigr) = \tilde\chi(L\tau)\, \chi(\tau)^{N-L} \ ,
\end{equation}
where $\tilde\chi$ is the NS sector character with the insertion of $(-1)^F$ --- this arises because under the exchange by an even cycle permutation fermionic states pick up a sign relative to bosonic states. The S-modular transformation then is of R-sector type, i.e., the leading exponent is 
\be
q^{\frac{1}{24 L}} \, q^{-\frac{N-L}{48}} \ .
\ee
Thus for $L$ even we get instead of (\ref{h0Lodd})
\be
h_0 = B\, \frac{L^2-1}{24 L} + F \, \frac{L^2+2}{48 L} 
\qquad \hbox{$L$ even} \ . 
\ee
For example, for the case of $\mathbb{T}^4$ for which $B=F=4$, this leads to 
\be\label{LevenT4}
h_0 = \frac{L}{4} \ , 
\ee
c.f.,  eq.~(2.20) of \cite{Lunin:2001pw}. 
However, for $F=8$ and $B=2$ we find instead 
\be\label{Levenground}
h_0 = \frac{L^2+1}{4L} \qquad \hbox{($L$ even)}  \ .
\ee

\end{document}